\pdfoutput=1
\documentclass[graybox]{svmult}

\usepackage{mathptmx}        
\usepackage{helvet}          
\usepackage{courier}         
\usepackage{type1cm}         
\usepackage{makeidx}         
\usepackage{graphicx}        
\graphicspath{{images/}{generated/}} 
\usepackage{multicol}        
\usepackage[bottom]{footmisc}

\usepackage{url}
\usepackage{cite}
\usepackage{array}

\makeindex             

\begin{document}

\title*{Using Provenance\\
  to support Good Laboratory Practice\\
  in Grid Environments}
\titlerunning{Using Provenance to support Good Laboratory Practice in
  Grid Environments}
\author{Miriam Ney\inst{1}
  \and Guy K. Kloss\inst{2}
  \and Andreas Schreiber\inst{1}}

\institute{Miriam Ney \and Andreas Schreiber
  \at Simulation and Software Technology,
  German Aerospace Centre,\\
  {Berlin, Cologne}, Germany\\
  \email{NeyMiriam@googlemail.com, Andreas.Schreiber@dlr.de}
  \and Guy K. Kloss
  \at School of Computing + Mathematical Sciences,
  Auckland University of Technology,\\
  Auckland, New Zealand\\
  \email{Guy.Kloss@aut.ac.nz}%
}

\maketitle

\abstract{Conducting experiments and documenting results is daily
  business of scientists.  Good and traceable documentation enables
  other scientists to confirm procedures and results for increased
  credibility.  Documentation and scientific conduct are regulated and
  termed as ``good laboratory practice.''  Laboratory notebooks are
  used to record each step in conducting an experiment and processing
  data.  Originally, these notebooks were paper based.  Due to
  computerised research systems, acquired data became more elaborate,
  thus increasing the need for electronic notebooks with data storage,
  computational features and reliable electronic documentation.  As a
  new approach to this, a scientific data management system
  (DataFinder) is enhanced with features for traceable documentation.
  Provenance recording is used to meet requirements of traceability,
  and this information can later be queried for further analysis.
  DataFinder has further important features for scientific
  documentation: It employs a heterogeneous and distributed data
  storage concept.  This enables access to different types of data
  storage systems (e.\,g.\ Grid data infrastructure, file servers).
  In this chapter we describe a number of building blocks that are
  available or close to finished development.  These components are
  intended for assembling an electronic laboratory notebook for use in
  Grid environments, while retaining maximal flexibility on usage
  scenarios as well as maximal compatibility overlap towards each
  other.  Through the usage of such a system, provenance can
  successfully be used to trace the scientific workflow of
  preparation, execution, evaluation, interpretation and archiving of
  research data.  The reliability of research results increases and
  the research process remains transparent to remote research
  partners.}

\section{Introduction}
\label{sect:Introduction}

With the ``Principles of Good Laboratory Practice and Compliance
Monitoring'' the OECD provides research institutes with guidelines and
a framework to ensure good and reliable research.  It defines ``Good
Laboratory Practice'' as \emph{``a quality system concerned with the
organisational process and the conditions under which non-clinical
health and environmental safety studies are planned, performed,
monitored, recorded, archived and reported''}~(p.~14
in~\cite{OECD_GLP_PrinciplesNo1}).  This definition can be extended to
other fields of research.  To prove the quality of research is of
relevance for credibility and reliability in the research community.
Next to organisational processes and environmental guidelines, part of
the good laboratory practice is to maintain a laboratory notebook when
conducting experiments.

The scientist documents each step, either taken in the experiment or
afterwards when processing data.  Due to computerised research
systems, acquired data increases in volume and becomes more elaborate.
This increases the need to migrate from originally paper-based to
electronic notebooks with data storage, computational features and
reliable electronic documentation.  For these purposes suitable data
management systems for scientific data are available.

\subsection{A Sample Use Case}
\label{sect:SampleUseCase}

As an example use case a group of biologists are conducting research.
This task includes the collection of specimen samples in the field.
Such samples may need to be archived physically.  The information on
these samples must be present within the laboratory system to refer to
it from further related entries.  Information regarding these samples
possibly includes the archival location, information on name, type,
date of sampling, etc.

The samples form the basis for further studies in the biological (wet)
laboratories.  Researchers in these environments are commonly not
computer scientists, but biologists who just ``want to get their
research done.''  An electronic laboratory notebook application
therefore must be similarly easy to operate in day-to-day practice
like a paper-based notebook.  All notes regarding experimentation on
the samples and further derivative stages (processing, treatments,
etc.) must be recorded, and linked to a number of other artifacts
(other specimen, laboratory equipment, substances, etc.).

As a result of this experimentation further artifacts are derived,
which need to be managed.  These could be either further physical
samples, or information (data, measurements, digital images,
instrument readings, etc.).  Along with these artifacts the team
manages documents outlining the project plan, documents on
experimental procedures, etc.

In the end every managed artifact (physical or data) must be linked
through a contiguous, unbroken chain of records, the provenance trail.
The biologists in our sample use case cooperate with researchers from
different institutes in different (geographical) locations.
Therefore, the management of all data as well as provenance must be
enabled in distributed environments, physically linked through the
Internet.  The teams rely on a common Grid-based authentication, which
is used to authorise principals (users, equipment, services) across
organisational boundaries.

The recorded provenance of all managed artifacts can be used in a
variety of ways.  Firstly, it is useful to document and \emph{prove}
proper scientific procedures and conduct.  Beyond this compliance
requirement provenance information can be used in further ways: It
enables often previously not possible (or very tedious) ways of
analysis.  By querying the present provenance information, questions
can be answered which depend on the recorded information.  These
questions may include some of the following:

\begin{itemize}
  \item \emph{Question for origin:} What artifacts were used in the
  generation of another artifact?

  \item \emph{Question for inheritance:} What artifacts and
  information were generated using a given artifact?

  \item \emph{Question for participants:} What actors (people,
  devices, applications, versions of tools, etc.) were employed in the
  generation of an artifact?

  \item \emph{Question for dependencies:} Which resources from other
  projects/processes have been used in the generation of an artifact?

  \item \emph{Question for progress:} In what stage of a processing
  chain is a given artifact?  Has the process the artifact is part of
  been finalised?

  \item \emph{Question for quality:} Did the process the artifact is
  part of reach a satisfactory conclusion by some given regulations or
  criteria?
\end{itemize}

\subsection{Data Management with the DataFinder}
\label{sect:DataFinder}

In order find a solution to common data management problems, the
German Aerospace Centre (DLR) -- as Germany's largest research
institute -- developed an open source data management application
aimed at researchers and engineers:
\emph{DataFinder}~\cite{SchlauchSchreiber2007_DataFinder--Scientific,
  DataFinderProject}.  DataFinder is a distributed data management
system.  It allows heterogeneous storage back-ends, meta-data
management, flexible extensions to the user interface and script-based
automation.  To implement required features for reliable and auditable
electronic documentation provenance technologies can be
used~\cite{BunemanKhannaEtAl2001_WhyandWhereDataProvenance}.

When analysing the data management situation in scientific or research
labs, several problems are noticeable:

\begin{itemize}
  \item Each scientist individually is solely responsible for the data
  generated and managing it as deemed fit.  Often others cannot access
  it, and duplication of effort may occur.

  \item If a scientist leaves the organisation, it is possible that no
  one understands the structure of the data left behind.  Information
  can be lost.

  \item Researchers often spend a lot of time searching for data.
  This waste of time decreases productivity.

  \item Due to long archiving periods and an increasing data
  production rate, the data volume to store increases significantly.
\end{itemize}

To overcome this situation common in many research institutes, the DLR
facility Simulation and Software Technology has developed the
scientific data management system DataFinder
(cf.~\cite{DataFinderProject}).

\subsubsection{General Concepts}

\begin{figure}[!h]
  \begin{center}
    \includegraphics[width=\textwidth]{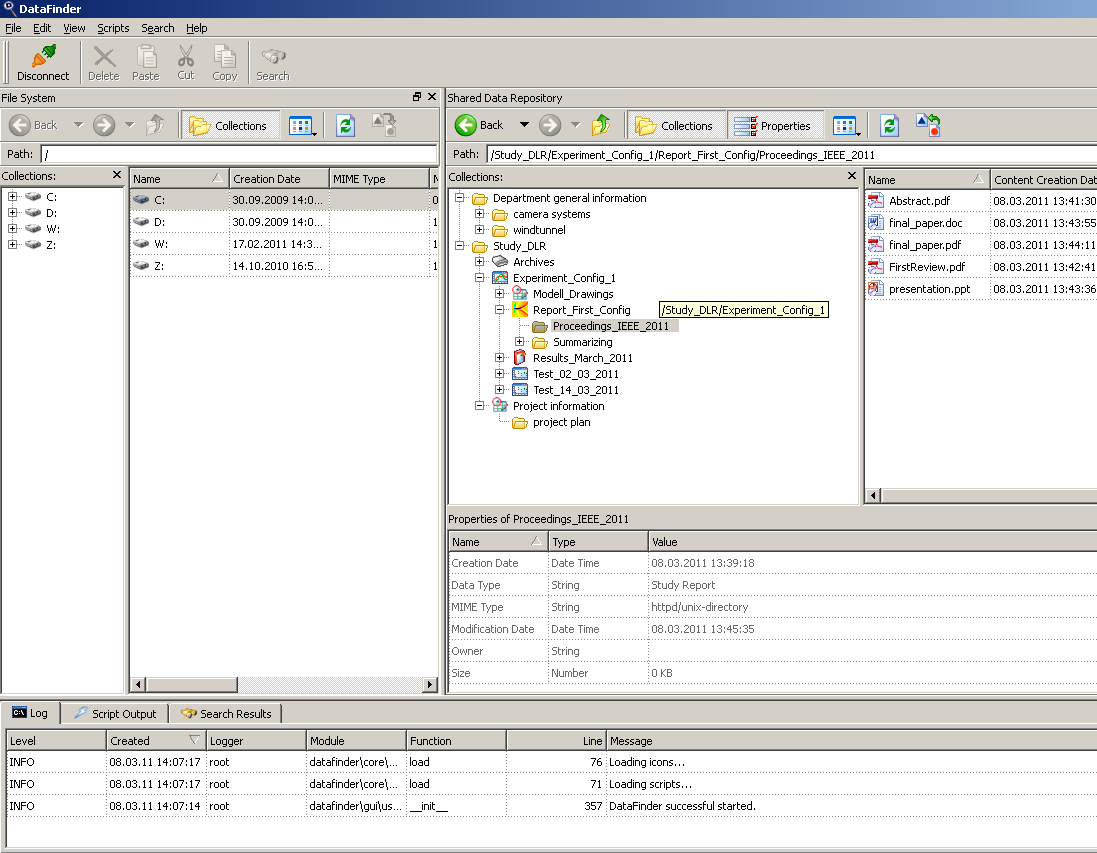}
    \caption{User interface of the DataFinder.}
    \label{fig:DFuserGUI}
  \end{center}
\end{figure}

DataFinder is an open source software written in Python.  It uses a
server and a client component.  The server component holds data and
associated meta-data.  Data and meta-data is aggregated in a shared
data repository and accessed and managed through the client
application.  Fig.~\ref{fig:DFuserGUI} shows the user interface of the
DataFinder, when connected to a shared repository.

It is designed similar to a file manager on common operating systems.
The left hand side presents the local file hierarchy, and the right
displays the shared repository.  All data on the server can be
augmented with arbitrary meta-data.  Common actions available for both
sides are: open, copy, paste, import and export data.  Opening an
entry will make an attempt to use the local system's default
association for a file.  These operations are all essential due to the
nature of DataFinder being a data management tool.  One must be aware
that on some operations (e.\,g.\ copying) provenance related
information is not copied with it.  Copying would create a fork in the
provenance graph to create a duplicate of a formerly uniquely
referenced artifact.  Special treatment to treat these cases in a way
as to extend the graph properly are not in place, yet.

An advantage of DataFinder is, that an individual data model is
configured for a shared repository, which must be followed by all its
users.  A data model defines the structure of collections.
Collections can contain (configurable) allowed data types, that can be
inserted into the collection.  The data model also defines a
pre-defined meta-data structure for these collections.  This meta-data
can be specified to be either optional or mandatory information when
importing a data item.  Based on the data model, data can be managed
on a heterogeneous storage system (certain data items stored in
different storage sub-systems, see~\ref{sect:DistributedDataStorage}).
This requires that DataFinder provides the ability to manage data on
different storage systems, under the control of a single user
interface under a single view (even within the same collection).

Lastly, it is the possibility to extend the application with Python
scripts.  This enables a user to take advantage of more customised
features, such as tool integration, task automation, etc.

The DataFinder-based system aims at providing many options and to be
highly extensible for many purposes.  DataFinder is already in use in
different fields of research.  New use cases are identified and
extensions implemented frequently.  One of these is the new use case
for supporting a good laboratory practice capable notebook as outlined
in this chapter.

\subsubsection{Distributed Data Storage}
\label{sect:DistributedDataStorage}

One of the key features of DataFinder is the capability to use
different distributed, heterogeneous storage systems (concurrently).
A user has the freedom to store data on different systems, while
meta-data for this data can be kept either on the same or on a
different storage system.

Possible data storage options can be accessed for example through:
WebDAV, Subversion, FTP, GridFTP\@.  Other available storage systems
possibilities are Amazon S3 Cloud services as well as a variety of
hosted file systems.  Meta-data for systems not capable of providing
extensive free-form meta-data is managed centrally with another
system.  Such systems then are accessed through meta-data capable
protocols like WebDAV or Subversion.  Further storage back-ends are
relatively simple to integrate, due to the highly modular factory
design of the application.  This design feature of DataFinder will be
further examined in
Sect.~\ref{sect:DistributedScientificDataManagement} for the
integration of a distributed Grid data storage infrastructure.

It must be noted at this point however, that DataFinder is responsible
for maintaining consistently managed data.  DataFinder uses these
protocols and systems for this purpose.  If data is accessed
\emph{without} using the DataFinder directly on the server through
other clients, data policies may be compromised (due to different
access restrictions), or consistency may be compromised (with writing
access to the storage systems).  With certain caution, this can
however be used to integrate other (legacy) systems into the overall
concept.

Due to the design of the DataFinder it is further possible to manage
physical (real world) items, such as laboratory analysis samples or
offline media (e.\,g.\ video tapes, CDs, DVDs).  Physical items can be
stored on shelves, or archived in any other way.  These can be
valuable artifacts for research, and the knowledge of their existence
as well as their proper management is a common necessity.  Therefore,
it is crucial to managed them electronically in a similar fashion by
the same management tools.  Doing so enables extensive meta-queries
provided by the DataFinder, taking advantage of utilising the search
capabilities over all managed items in the same way.  Furthermore,
this enables to reference them consistently in provenance assertions
from within the realm of the provenance enabled system.

\subsection{Overview}

This chapter ties the link between the existing DataFinder application
to convert it into a tool useful for a good laboratory practice
compliant electronic notebook.  It will introduce how DataFinder can
be combined with provenance recording services and (Grid) storage
servers to form the back bone of such a system.  The concept of
DataFinder is to be a system that can be customised towards different
deployment scenarios, it is to support the researchers or engineers in
\emph{their} way of working.  This includes the definition of a data
storage hierarchy, required meta-data for storage items and much more,
usually alongside with customisation or automation scripts and
customised GUI dialogues.  In a similar fashion, DataFinder can be
used to construct an electronic laboratory notebook with provenance
recording for good laboratory practice.  Again, to do so one creates
the required data models and customises GUI dialogues to suit the
purpose.
  
The used provenance technologies and their applications are described
in Sect.~\ref{sect:ProvenanceManagement}.  Concepts to integrate Grid
technologies for scientific data management are outlined in
Sect.~\ref{sect:DistributedScientificDataManagement}.
Sect.~\ref{sect:Results} presents the results of integrating the good
laboratory practice into a provenance system as well as a data
management system.  It also provides a solution on how to connect
these two system practically.  Finally, the concept of the resulting
system of an electronic laboratory notebook is evaluated.


\section{Provenance Management}
\label{sect:ProvenanceManagement}

Provenance originates from the Latin word: ``provenire'' meaning ``to
come from''~\cite{Merriam-Webster2010_OnlineDictionary}.  It is
described as ``the place that sth.\ originally came from'' thus the
origin or source of something
(cf.~\cite{Wehmeier2000_OxfordAdvancedLearnersDict}).  It was
originally used for art, but other disciplines adapted it for their
objects, such as fossils or documents.  In the field of computer
science and data origin it could be defined as:

\begin{quote}
  ``The provenance of a piece of data is the process that led to that
  piece of data.''~\cite{Moreau2010_FoundationsProvenanceWeb}
\end{quote}

\begin{figure}[!h]
  \begin{center}
    \includegraphics[width=\textwidth]{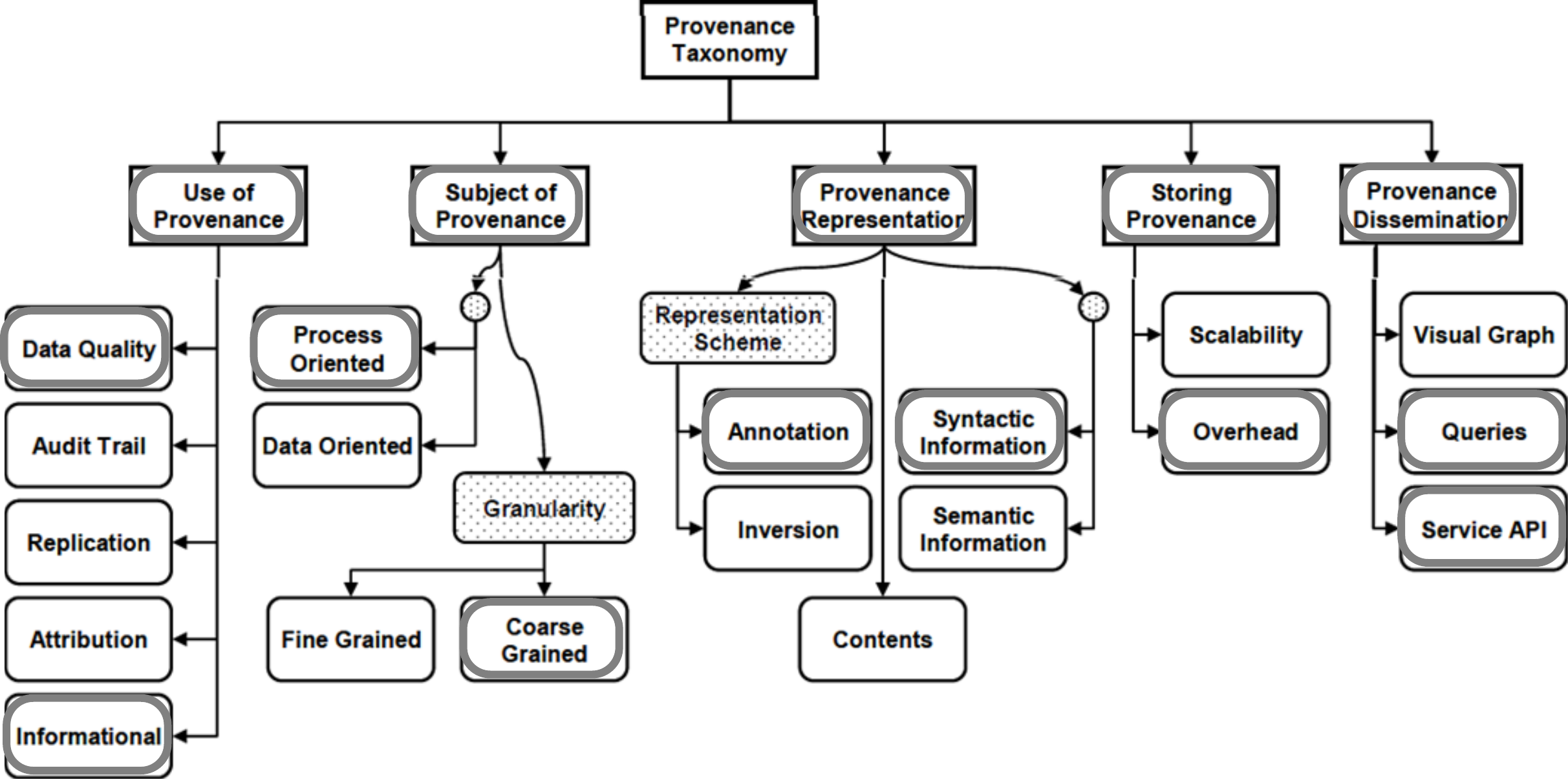}
    \caption{Provenance taxonomy according
      to~\cite{SimmhanPlaleEtAl2005_SurveyDataProvTech}.}
    \label{fig:prov_tax}
  \end{center}
\end{figure}

Based on this understanding, approaches for identifying provenance use
cases for modeling processes and for integrating provenance tracking
into applications are developed.  Also, concepts to store and
visualise provenance information are investigated.  An overview of the
different areas of provenance gives Fig.~\ref{fig:prov_tax}.

The figure shows five major areas: \emph{Usage, Subject,
  Representation, Storage and Dissemination.}
\cite{SimmhanPlaleEtAl2005_SurveyDataProvTech}~gives a detailed
description on each area and their subdivisions.  In this application,
embedded provenance tracking in the data management system enables
DataFinder to provide information about the chain of steps or events
leading to a data item as it is.  The following list outlines relevant
elements of the taxonomy from Fig.~\ref{fig:prov_tax} (additionally
framed elements):

\begin{description} 
  \item[Use of provenance:] Provenance is used to present
  \emph{information} of the origin of the data, but also to provide
  \emph{data quality.}

  \item[Subject of Provenance:] The subject is the \emph{process} of
  conducting a study or experiment.  It is focused on documentation.
  To identify the subject further, the Provenance Incorporating
  Methodology (PrIMe, Sect.~\ref{sect:ProvenanceModelForGLP}) is used.

  \item[Provenance Representation:] Provenance information will be
  represented in an \emph{annotational} model, based on the Open
  Provenance Model (OPM, Sect.~\ref{sect:OPM}) and it will mainly hold
  \emph{syntactic information.}

  \item[Storing Provenance:] Provenance information will be stored in
  the prOOst (Sect.~\ref{sect:prOOst}) system (can also hold
  additional information).

  \item[Provenance Dissemination:] To extract provenance information,
  the provenance system can be queried using a graph traversal
  language (Sect.~\ref{sect:Gremlin}).
\end{description}

The main concepts of OPM and the provenance system prOOst are
described in the following sections, whereas PrIMe is discussed in the
scope of applying the technical system to the domain of good
laboratory practice in Sect.~\ref{sect:Results}.

\subsection{OPM -- Open Provenance Model}
\label{sect:OPM}

The Open Provenance
Model~\cite{MoreauCliffordEtAl2010_OpenProvenanceModel} is the result
of the third ``provenance challenge''
efforts~\cite{SimmhanGrothEtAl2011_ThirdProvChallengeOPM} to provide
an interchangeable format between provenance systems.  In its core
specification, it defines elements (nodes and edges) to describe the
provenance of a process.

%

\begin{figure}[!h]
  \begin{center}
    \includegraphics[width=0.8\textwidth]{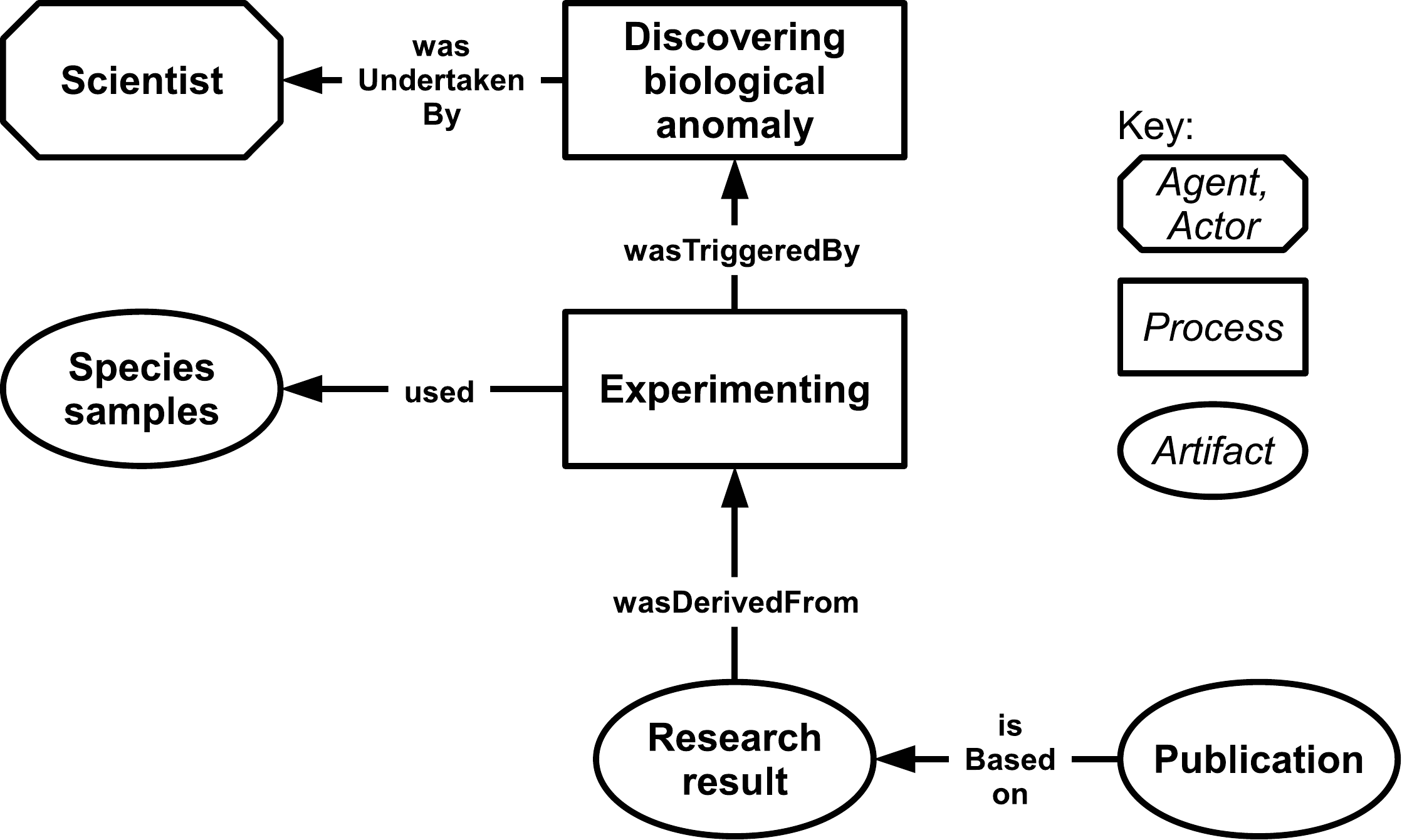}
    \caption{Example a biological study as an OPM model.}
    \label{fig:opm_example}
  \end{center}
\end{figure}

Nodes can be \emph{processes, agents/actors} and \emph{artifacts/data
  items.}  The nodes can be connected through edges, such as
\emph{``used'', ``wasUndertakenBy'', ``wasTriggeredBy'',
  ``wasDerivedFrom''} and \emph{``isBasedOn''.}  Each edge is
directed, clearly defining the possible relations within a provenance
model.  Each node can be enriched by annotations.
Fig.~\ref{fig:opm_example} gives an example for conducting experiments
in a biological laboratory and it shows the usage of the model
notation.  In the example, a scientist (actor) discovers a biological
anomaly (controls the process of thinking and inspiration).  So he
starts experimenting (triggered by the discovery).  For it to produce
research results (derived from experimenting), he needs (uses)
specimen samples to work on.  If the results show a significant
research outcome, a research paper can be written (based on) the
results.

\subsection{Provenance Storage with prOOst}
\label{sect:prOOst}

Groth et al.\ describe
in~\cite{GrothMilesEtAl2005_ArchitectureProvenanceSystems}
theoretically the architecture of a provenance system.
In~\cite{MoreauCliffordEtAl2010_OpenProvenanceModela} the
representation of a provenance system is described as follows: A
provenance aware application sends information of interest to the
provenance store.  From this store inquiries and information is
gathered, and possibly given back to the application.

To record the information, different approaches have been
investigated.  In~\cite{HollandBraunEtAl2008_ChoosingDataModel} four
different realisations are discussed: Relational, XML with XPath, RDF
with SPARQL and semi-structured approaches.  They conclude
semi-structured approaches to be most promising.  In semi-structured
systems, the used technology has no formal structure, but it provides
means of being queried.

This work uses a semi-structured approach for the provenance storage
system \emph{prOOst.}  It uses the graph database
``Neo4j''~\cite{Neo4jProject} for storage and the graph traversal
language ``Gremlin''~\cite{GremlinProject} for querying.  Furthermore,
it provides a REST interface to record data into the store, and a web
front end to query the database.  The prOOst provenance system was
published under the Apache license in July 2011 on
SourceForge.\footnote{\url{http://sourceforge.net/projects/proost/}}

It is not the first implementation using a graph database for storage
technology.
In~\cite{TylissanakisCotronis2009_DataProvenanceReproducibility} this
approach was already successfully tested.  Neo4j was chosen as it is a
robust, performant and popular choice for graph storage systems.
Additionally it readily connectible with the suitable Gremlin query
system to meet our requirements.  Further discussions on alternative
storage or query systems are outside the scope of this chapter.
Further information on the implementation of OPM model provenance
assertions using these systems are described in the following two
sections.

\subsubsection{Graph Database: Neo4j}

\begin{quote}
  ``Neo4j is a graph database, a fully transactional database that
  stores data structured as graphs.'' (cf.~\cite{Neo4jProject})
\end{quote}

An advantage of graph databases like Neo4j is that they offer very
flexible storage models, allowing for a rapid development.  Neo4j is
dually licensed (AGPLv3 open source and commercial).

\begin{figure}[!h]
  \begin{center}
    \includegraphics[width=\textwidth]{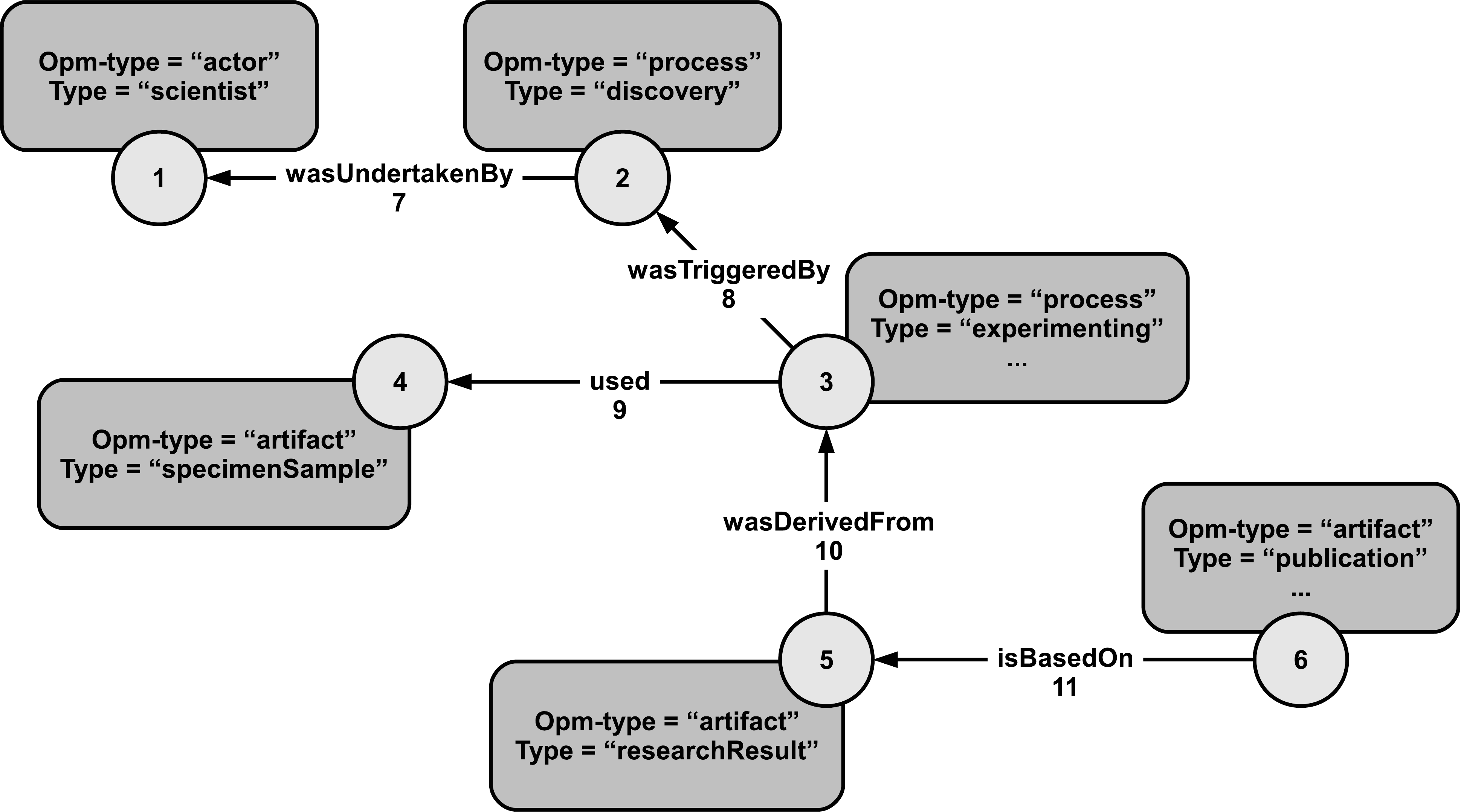}
    \caption{OPM example in Neo4j.}
    \label{fig:OPM_Neo4j}
  \end{center}
\end{figure}

Modelling OPM using Neo4j is described in more detail
in~\cite{Wendel2010_UsingProvenanceSoftwareDev}.
Fig.~\ref{fig:OPM_Neo4j} shows the previous example (from
Fig.~\ref{fig:opm_example}) modelled as an OPM graph.  Each element is
represented by a node (vertex) in the database.  Nodes are indexed
according to the Neo4j standard.  The nodes can be annotated with
further (OPM specific) information, such as ``process'' or
``artifact''.  Analogously, also the edges connecting the nodes are
indexed and annotated with a label (the OPM relationship).

\subsubsection{Query Language: Gremlin}
\label{sect:Gremlin}

``Gremlin is a graph traversal language''~\cite{GremlinProject}.
Gremlin already provides an interface to interact with the Neo4j graph
database.  The following example shows its use for querying Neo4j on
the example database, searching for the names (identifiers) of all
discoveries of a certain \texttt{scientistX}:

\begin{verbatim}
$_g := neo4j:open('database')
$scientists := g:key($_g, 'type', 'scientist')
$scientistX := g:key($scientists, 'identifier', 'scientistX')
$discoveries := $scientistX/inE/inV[@identifier']
\end{verbatim}


\section{Distributed, Scientific Data Management}
\label{sect:DistributedScientificDataManagement}

The previous sections have discussed the technical means to manage
data on the user side (DataFinder) and to store and query the
provenance information.  As indicated, DataFinder can handle a variety
of different data storage servers.  However, to store data and its
associated meta-data on the same system, and to take full advantage of
Grid technologies for cross-organisational federated access, a
suitable data storage service has to be chosen.  For the example use
case of the team of biologists, federated access management (e.\,g.\
through Shibboleth\footnote{\url{http://shibboleth.internet2.edu/}})
and integration with further Grid-based resources would be desired
(e.\,g.\ for resources to compute on sequenced genome data).

An electronic laboratory notebook system is a data management system,
only with the particular needs towards managing the experimental and
laboratory relevant data in a suitable fashion.  This can generally be
accomplished by tweaking a generic storage system for data and
(extensive) associated meta-data towards the use case for supporting
good laboratory practice.  This section therefore mainly raises the
questions towards the use of such storage systems in Grid-based
environments.

Various ways are possible to envision for making relevant data
available to researchers in distributed teams.  Commonly encountered
mechanisms in such (Grid) research environments are based on top of
GridFTP (the ``classic'' Grid data protocol) or WebDAV (extension to
the HTTP protocol).  In some environments more full featured
infrastructures, like iRODS\footnote{\url{http://www.irods.org/}} have
been deployed.  One such environment is the New Zealand based ``Data
Fabric'' -- as implemented for the New Zealand eScience Infrastructure
(during the recently concluded BeSTGRID project).  iRODS offers data
replication over multiple geographically distributed storage
locations, with one centralised meta-data catalogue.  Its data is
exposed through the iRODS native tools and libraries, as well as
through WebDAV (using Davis\footnote{WebDAV-iRODS/SRB gateway:
  \url{http://projects.arcs.org.au/trac/davis/}}), a web-based
front-end and GridFTP (through the Griffin GridFTP
server~\cite{ZhangCoddingtonEtAl2010_GriffinForArbitraryDataSources,
  ZhangKlossEtAl_GriffinProject} with an iRODS back-end using
Jargon\footnote{\url{https://www.irods.org/index.php/Jargon}}).

\subsection{Integration of Existing Storage Servers}
\label{sect:IntegratingExistingStorageServers}

We are discussing data integration solutions according to the above
mentioned scenario of the New Zealand Data Fabric.  From this, slight
variations of the setup can be easily extrapolated.

Three obvious possibilities exist to use this type of infrastructure
for provenance enabled data management and/or as a laboratory notebook
system for distributed environments.  For all these, users need to be
managed and mapped between multiple systems, as iRODS introduces its
own mandatory user management.  This may only be required for the
storage layer, but it does introduce a redundancy.  The options are
discussed in the following paragraphs.

The easiest, and directly usable, way is to integrate this Grid Data
Fabric as an external \emph{WebDAV} data store, using the existing
persistence module.  Even though WebDAV is a comprehensive storage
solution for the DataFinder for data and meta-data, this service layer
on top of iRODS does not permit the required WebDAV protocol means to
access the meta-data.  An additional meta-data server is required, and
therefore potentially multiple incompatible and separate sets
meta-data may exist for the same data item stored.  Unfortunately this
WebDAV service does not use the full common Grid credentials for
access, but is limited to MyProxy\footnote{Software for managing X.509
  Public Key Infrastructure (PKI) security credentials:
  \url{http://grid.ncsa.illinois.edu/myproxy/}} based authentication
as a work around.

As the next step up, DataFinder can be equipped with a \emph{GridFTP}
back-end in its persistence layer.  Such a module was already
available for a previous version (1.3) of DataFinder, and only
requires some porting effort for the current (2.x) series.  Again,
GridFTP is only able to access the payload data, and is not capable to
access any relevant meta-data, resulting in the need of an additional
and separate meta-data service.  An advantage is that this solution
uses the common Grid credentials for authentication.

Lastly, the development of a native \emph{iRODS} storage back-end
based on the txIRODS Python
bindings\footnote{\url{http://code.arcs.org.au/gitorious/txirods}} is
a possibility.  This solution could also use the iRODS meta-data
capabilities for native storage on top of the payload data storage.
Unfortunately, this last solution also requires the use of the native
iRODS user credentials for accessing the repository, as it is
completely incompatible to any of the common Grid authentication
procedures.

The above mentioned scenarios can be freely modified, particularly the
first two regarding their underlying storage infrastructure.  One
could deploy other storage systems that expose access using WebDAV or
GridFTP as service front ends for simplicity, potentially sacrificing
any of the other desired features of iRODS like cross-site
replication.

When sketching out a potential deployment, the above mentioned
scenarios did not strike us as being particularly nice to implement or
manage.  Several shortcomings were quite obvious.  Firstly, the
central meta-data catalogue, which can turn out to be a bottle neck.
Particularly meta-data heavy scenarios requiring extensive queries on
meta-data would suffer due to increased latencies.  Secondly, the
iRODS system provides a multitude of features, which make the system
implementation as well as its deployment at times quite convoluted.  A
simpler, more straight forward system is often preferred.  Lastly,
multiple user management systems can be an issue, particularly if this
includes the burden of mapping between, particularly if they are based
on different concepts.  Grid user management is conceptually based on
cross-organisational federation, including virtual organisations (VOs)
and delegation using proxy certificates, which cannot be neatly
projected to other user concepts as employed by iRODS.

\subsection{Designing an Alternative Storage Concept -- MataNui}
\label{sect:MataNui}

The idea for an alternative storage solution came up, which is simpler
and a better ``Grid citizen.''  For performant storage of many or
large files inclusive meta-data, the NoSQL database MongoDB with its
driver side file system implementation ``GridFS'' seemed like a good
choice.  A big advantage of this storage concept is, that MongoDB can
perform sharding (horizontal partitioning) and replication
(decentralised storage with cross-site synchronisation) ``out of the
box.''  Therefore, the only concerns to target were to provide
suitable service front-ends to the storage sub-system, to offer the
capabilities for the required protocols and interfaces to the
DataFinder.  This means that research teams can opt for running local
server instances (alternatively to accessing a remote server) for an
increase in performance as well as decrease in latencies.  This local
storage sub-system also increases data storage redundancy, which leads
to a better fault protection in cases of server or networking
problems.  Each storage server individually can be exposed through
different service front-ends, reducing bottle necks.  These service
front-ends can be deployed in a site specific manner, reducing the
number of server instances to those required for a site.

This distributed storage concept for data and meta-data, complimented
by individual front-end services in a building block fashion, has been
dubbed
``MataNui''~\cite{Kloss2010_MataNuiBuildingGridDataInfrastructure}.
The MataNui server~\cite{Kloss_MataNuiProject} itself provides full
access to all content, including server side query capability and
protection through native Grid (proxy) certificate authentication
(X.509 certificates).  As the authentication is based on native Grid
means, it is obvious to base the user management on Grid identities as
well, the distinguished names (DN) of the users.  MataNui is based on
a REST principle based Web Service (using JSON encoding), and is
therefore easy to access through client side implementations.

Exposing further server side protocols is done by deploying generic
servers, that have been equipped with a storage back-end accessing the
MataNui data structures hosted in the MongoDB/GridFS containers.  It
was relatively simple to implement the GridFTP protocol server on the
basis of the free and open
Griffin~\cite{ZhangCoddingtonEtAl2010_GriffinForArbitraryDataSources,
  ZhangKlossEtAl_GriffinProject} server.  A first beta development
level GridFS back-end is already part of the Griffin code base.
Possibly later a WebDAV front end is going to be implemented,
equipping one of the quite full featured
Catacomb\footnote{\url{http://catacomb.tigris.org/}} or
LimeStone\footnote{\url{https://github.com/tolsen/limestone}} servers
with a GridFS back-end for data and meta-data.  Such servers then
could also be used to access (and query) the meta-data through the
WebDAV protocol, if the storage back-end supports this.  Lastly, it is
even possible to use a file system driver to mount a remote GridFS
into the local Linux/UNIX system.  However, access control to the
content is provided through the services on top of the MongoDB/GridFS
server.  Therefore, this will likely circumvent any protective
mechanisms.  A better solution would be to mount a WebDAV exposed
service into a local machine's file system hierarchy.

\begin{figure}[!h]
  \begin{center}
    \includegraphics[trim=8mm 28mm 0mm 8mm, clip,%
    width=\textwidth]{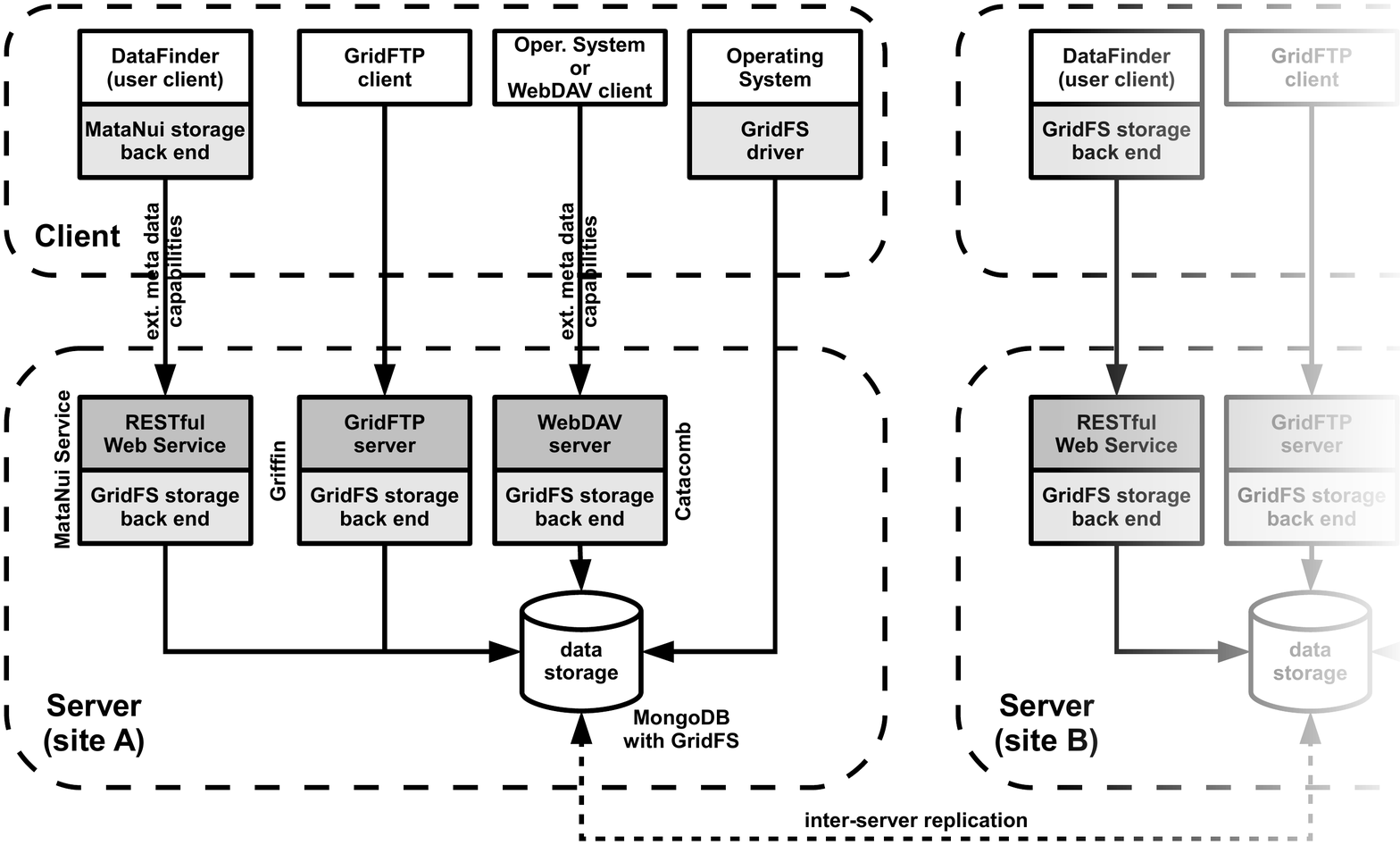}
    \caption{Conceptual links between components in a Grid-based data
      fabric to support researchers in distributed environments.  The
      system provides for decentralised access to geographically
      distributed data repositories, while enabling administrators to
      only expose local storage through service front-ends as
      required.}
    \label{fig:MataNui}
  \end{center}
\end{figure}

Access through protocols as GridFTP and WebDAV is quite straight
forward through various existing clients in day-to-day use within the
eResearch communities.  This is different with the MataNui RESTful
service.  As outlined in
Sect.~\ref{sect:IntegratingExistingStorageServers} already, the
DataFinder can be quite easily extended towards providing further
persistence back-ends, like a potential iRODS back-end.  In a similar
fashion a MataNui REST service client back-end can be implemented.
The big difference being, that it does not require any external
modules that are not well maintained.  It can mainly be based on the
already available standard library for HTTP(S) server access, with the
addition of suitable cryptography provider for extended X.509
certificate management.  This can be done either by simply wrapping
the OpenSSL command line tool or by using one of the mature and well
maintained libraries such as
pyOpenSSL.\footnote{\url{https://launchpad.net/pyopenssl}}
  
This modularity of service front-ends leaves administrators the option
to set up sites with exactly the features required locally.  However,
in a global perspective, MataNui enables a new perspective on the
functionality of a data fabric for eResearch.  Fig.~\ref{fig:MataNui}
provides a conceptual overview of how such a distributed data
repository can be structured.  Every storage site requires an instance
of MongoDB with GridFS\@.  These are linked with each other into a
replication set (with optional sharding).  The storage servers for the
different sites expose the repository through one or more locally
hosted services, such as the MataNui RESTful Web Service, a GridFTP
server, etc.  These services can be accessed by clients suitably
equipped for the particular service.  Clients, such as the DataFinder,
may require an additional implementation for a particular persistence
back-end.  Some of these clients (e.\,g.\ DataFinder or a WebDAV
client) may be equipped to take advantage of the full meta-data
capabilities of the data fabric, whereas others (e.\,g.\ GridFTP or
file system mounted WebDAV) may only access the data content along
with some rudimentary system meta-data (time stamp, size, etc.).

In a scenario like this data and its meta-data can be managed in the
distributed environment through DataFinder.  Seamless integration when
working with other Grid resources is unproblematic: All systems share
the same type of credentials, and data can be transferred between Grid
systems directly through GridFTP without the need of being routed
through the user's workstation.


\section{Results}
\label{sect:Results}

The following describes the application of the previously discussed
technologies to implement the provenance enabled electronic laboratory
notebook.  For this also the data management system DataFinder
requires customisation (through Python script extensions) to suit the
users' needs.  It is enhanced with features to trace documentation.

First the development of the provenance model for good laboratory
practice by means of the PrIMe methodology is described in
Sect.~\ref{sect:ProvenanceModelForGLP}.  Required modifications
applied in the DataFinder code are outlined in
Sect.~\ref{sect:DataFinderGLPAdjustments}.
Sect.~\ref{sect:IntegrationEvaluationOfLaboratoryNotebook} evaluates
the integration of DataFinder for the purpose of use as an electronic
laboratory notebook in a final system.  More information on this
evaluation can be found in~\cite{Ney2011_GLP}.  Lastly,
Sect.~\ref{sect:OutlookImprovingDataFinderNotebook} gives an outlook
on improving DataFinder in its role as an electronic laboratory
notebook, as well as on deploying such an infrastructure fully to Grid
environments.

\subsection{Developing a Provenance Model for Good Laboratory
  Practice}
\label{sect:ProvenanceModelForGLP}

Munroe et al.~\cite{MunroeMilesEtAl2006_PrIMe:MethodologyDeveloping}
developed the PrIMe methodology to identify parameters for
``provenance enabling'' applications.  These parameters then can be
used to answer provenance questions.  A provenance question usually
identifies a scenario, in which provenance information is needed.
Questions relevant for the analysis, are for example: ``Who inserted
data item $X$?'', ``What data items belong to a report $X$?'' and ``
What is the logical successor of data item $X$?''.

\begin{figure}[!h]
  \begin{center}
    \includegraphics[width=\textwidth]{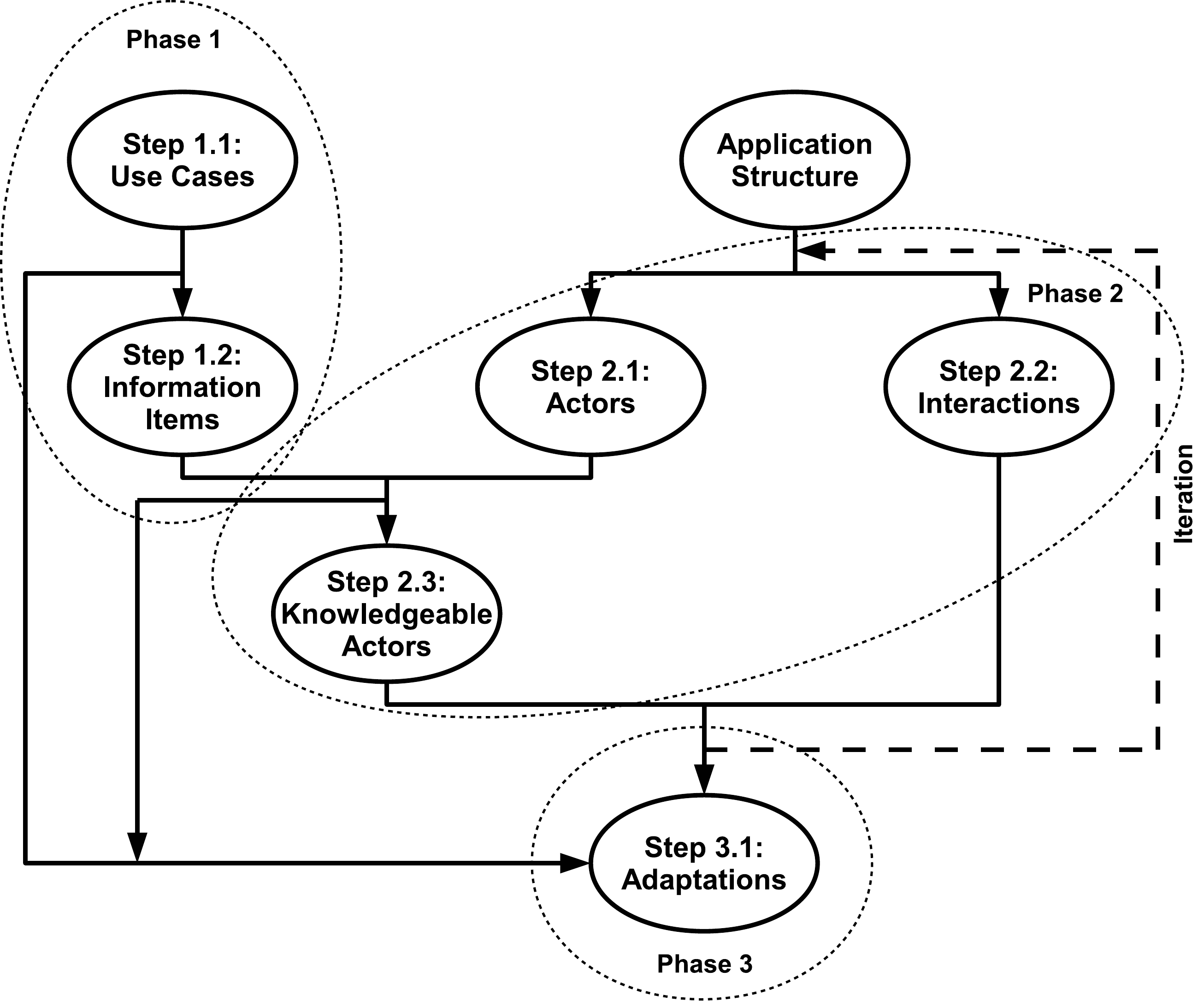}	
    \caption{Structure of PrIMe
      approach~\cite{MunroeMilesEtAl2006_PrIMe:MethodologyDeveloping}.}
    \label{fig:prime_structure}
  \end{center}
\end{figure}

This approach was modified
(in~\cite{Wendel2010_UsingProvenanceSoftwareDev}), as it used the
older p-assertion protocol (p.~15
in~\cite{Wendel2010_UsingProvenanceSoftwareDev} and p.~2
in~\cite{MunroeMilesEtAl2006_PrIMe:MethodologyDeveloping}) instead of
the now more common Open Provenance Model
(OPM)~\cite{MoreauCliffordEtAl2010_OpenProvenanceModel}.  The
p-assertion protocol is similar in use to OPM, so the approach can
easily be adapted.  The following list describes the three phases of
the adapted PrIMe version in correspondence to the PrIMe structure
from Fig.~\ref{fig:prime_structure}:

\begin{description}
  \item[Phase 1:] ``In phase~1 of PrIMe, the kinds of provenance
  related questions to be answered about the application must be
  identified''~\cite{MunroeMilesEtAl2006_PrIMe:MethodologyDeveloping}~(p.~7).
  First, provenance \emph{questions} are determined.  Then,
  corresponding \emph{data items/artifacts} that are relevant to the
  the answer, are investigated.

  \item[Phase 2:] \emph{Sub-processes, actors} and \emph{interactions}
  are identified in phase~2.  The sub-processes are part of the
  adaptation (Step~2.1).  Actors generate data items or control the
  process.  Relations between sub-processes and data items are defined
  as interactions (Step~2.2).  Actors, processes and interactions are
  modeled with OPM.

  \item[Phase 3:] The last phase finally adapts a system to the
  provenance model.  In this phase, the provenance store is populated
  with information from the application.  In the discussed scenario,
  this is accomplished via REST requests to the storing system.
\end{description}

Some exemplary questions that could be relevant in the sample use case
have already been given in Sect.~\ref{sect:SampleUseCase}.  After
analysing the questions, participating processes need to be
identified.  A scientific experiment for which documentation is
provided can be divided into five sub-processes:

\begin{enumerate}
  \item Preparation of the experiment, generating a study plan.

  \item Execution of the experiment according to a study plan,
  generating raw data.

  \item Evaluation of raw data, making them processable for
  interpretation.

  \item Interpretation of data, publishing it or processing the data
  further.

  \item Preservation of the data according to regularities.
\end{enumerate}

The very generic nature of these sub-processes is mandated by the OECD
principles of good laboratory practice~\cite{OECD_GLP_PrinciplesNo1}.
Obviously, researchers can augment each of these with further internal
sub-processes as required by the project or studies undertaken.

\begin{figure}[!h]
  \begin{center}
    \includegraphics[width=\textwidth]{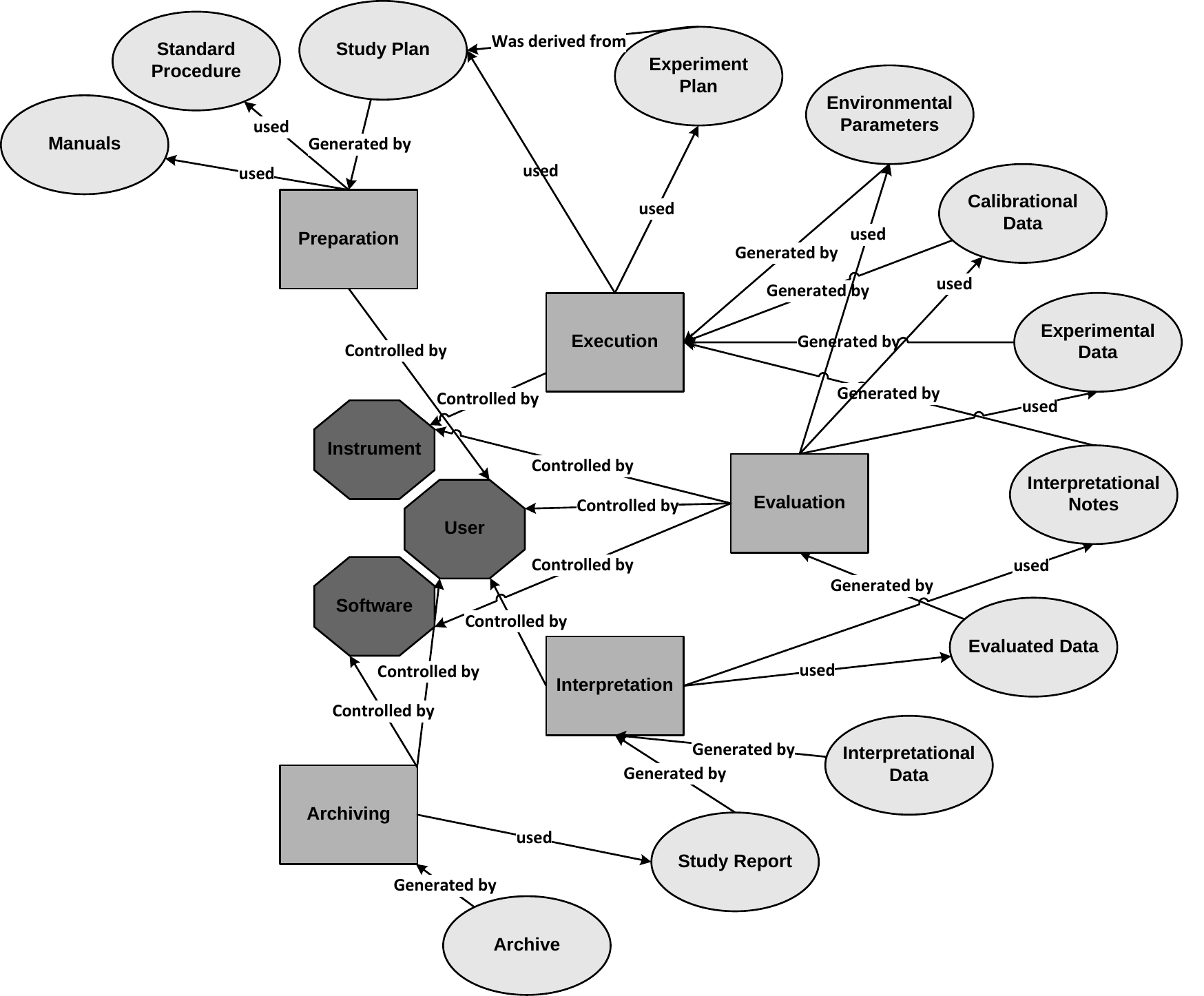}
    \caption{OPM for scientific experiment documentation.}
    \label{fig:OPM_all}
  \end{center}
\end{figure}

These sub-processes are modeled with the Open Provenance Model (OPM).
Fig.~\ref{fig:OPM_all} shows the model in OPM notation for good
laboratory practice.  The five rectangles in the figure symbolise the
above mentioned sub-processes.  Data items/artifacts are indicated by
circles: These are managed by the DataFinder.  Lastly, the octagons
represent the actors controlling the processes.

Provenance information is gathered in the data management system on
data import and modification.  Then -- according to the provenance
model -- this information is sent to a provenance storage system (as
described in Sect.~\ref{sect:ProvenanceManagement}).

\subsection{Adjustments for Good Laboratory Practice in the
  DataFinder}
\label{sect:DataFinderGLPAdjustments}

\begin{figure}[!h]
  \begin{center}
    \includegraphics[width=\textwidth]{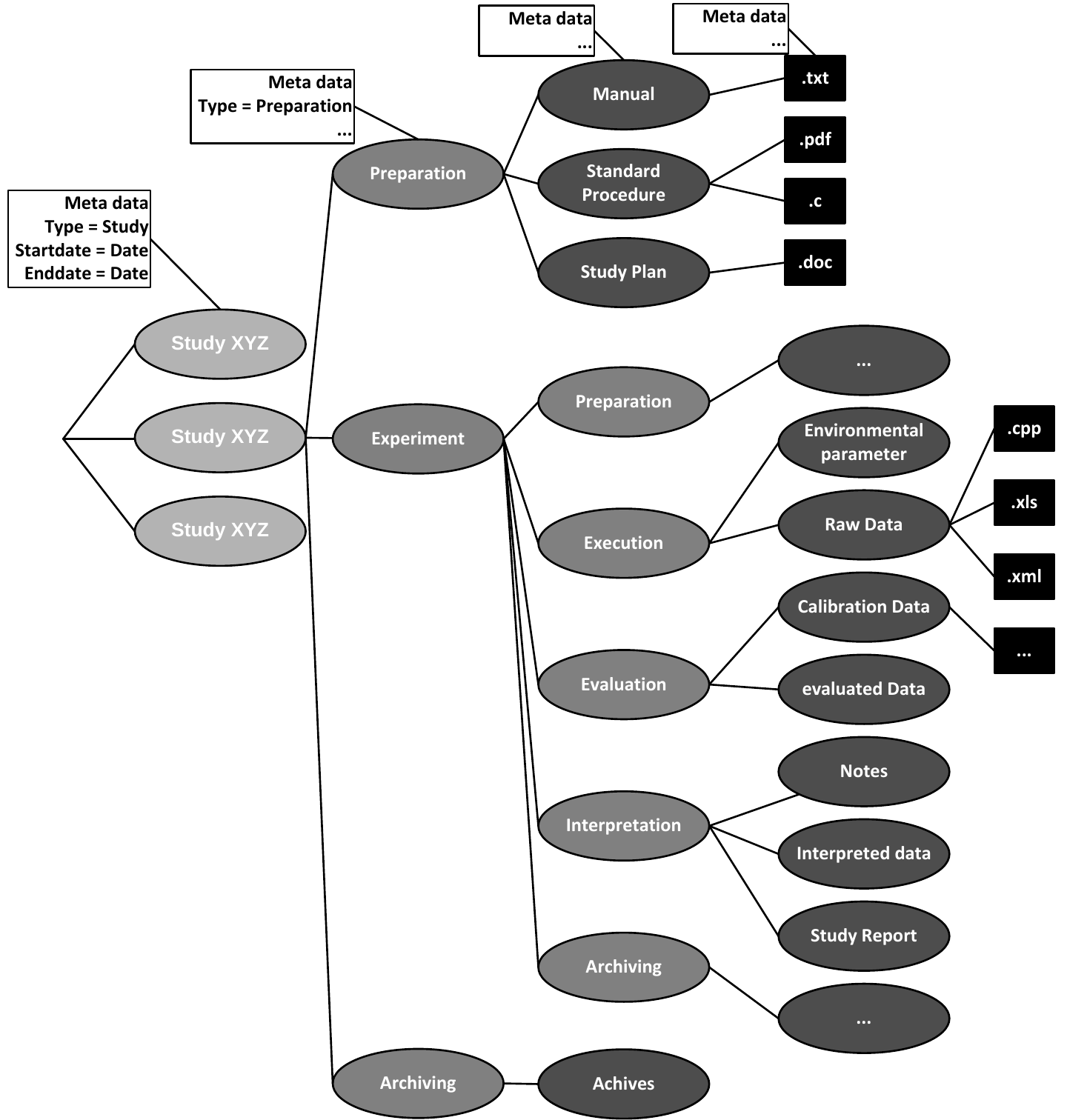}
    \caption{Laboratory notebook data model used for the DataFinder.}
    \label{fig:DFdatamodel}
  \end{center}
\end{figure}

To use the DataFinder as a supportive tool for good laboratory
practice, a new data model and Python extensions were developed.  The
main part of the data model is presented in
Fig.~\ref{fig:DFdatamodel}.

\label{sec:extensions}

The model is derived through requirements analysis
in~\cite{OECD_GLP_PrinciplesNo1}.  It divides the data into the five
major categories according to Sect.~\ref{sect:ProvenanceModelForGLP}:
Preparation, execution, evaluation, interpretation and archiving.  All
experiments pass through theses categories in their five processes.
Each process needs or generates different types of data.  Data is
aggregated in (nested) \emph{collections,} the data repository
equivalent of directories in a file system.  Collections representing
these processes aggregate data items belonging to that process.  Each
collection or element can mandate attached meta-data (such as type or
dates).  The data model also provides structural elements at a higher
level of the hierarchy to differentiate between different studies and
experiments.  Processes and data items are reflecting the model
structure in Fig.~\ref{fig:OPM_all}.  The DataFinder repository
structure is defined through its underlying data model\footnote{The
  complete data model is described in XML and available on
  \url{https://wiki.sistec.dlr.de/ DataFinderOpenSource/
    LaboratoryNotebook}} (implemented according to the OPM model).
 
In the screen shot of Fig.~\ref{fig:DFuserGUI} at the beginning of
this chapter, a user is connected to a shared repository (left side)
operating on the described data model.  The user now is required to
organise data accumulated according to this model.  For example, a new
collection of manuals may only be created within a parent collection
of the ``preparation'' type.  A ``preparation'' collection can then be
either part of a ``study'' or ``experiment.''

Three further extensions to DataFinder have been developed.  They are
needed to support good laboratory practice in DataFinder:

\begin{itemize}
  \item The most important extension is an \emph{observer mechanism,}
  listening on \emph{import events} into the DataFinder.  Upon the
  import of a new document, it reacts by prompting with a dialog
  asking for input items within the system that have influenced the
  data item/artifact.  After analysing the corresponding process, the
  information is recorded in the provenance store.

  \item A second extension supports \emph{evidential archiving.}  For
  this the user can send an archive to an archiving service, to
  analyses the credibility of the archive.\footnote{This is not
    further discussed, because it is a separate project in Germany.}
  To provide sufficient information, the user can activate a specific
  script extension, which generates an archive composed of information
  relevant to the data from information in the provenance store.  The
  user selects a study report, and the provenance store is queried for
  all data items influencing the report for each step.

  \item Lastly, a \emph{digital signing mechanism} was implemented
  through an extension, aiming at increasing credibility of data items
  through non-repudiation.
\end{itemize}

\subsection{Integration Evaluation of an Electronic Laboratory
  Notebook}
\label{sect:IntegrationEvaluationOfLaboratoryNotebook}

Tab.~\ref{tab:DF_req_lab} evaluates the DataFinder concepts on the
requirements defined in Chap.~3.1 of~\cite{Ney2011_GLP}.  It explains
how each requirement is integrated into the DataFinder
system.\footnote{The table and its description is adapted
  from~\cite{Ney2011_GLP}} The table shows that almost all
requirements are either already currently met, are implemented through
extensions as described here, or otherwise currently implemented.  As
a result, DataFinder can be used as laboratory notebook, supporting
the concepts of good laboratory practice, and is therefore supportive
to scientific working methods.

\begin{table}
  \begin{tabular}{|l>{\raggedright\arraybackslash}p{3cm}%
      |>{\raggedright\arraybackslash}p{3cm}%
      |>{\raggedright\arraybackslash}p{5cm}|}
    \hline
    \multicolumn{2}{|l|}{Requirement} & Implemented? & Details \\[1ex]
    \hline
    \hline
    \textbullet & Chain of events
    & yes\newline
    with extensions described here
    & provenance for modeling the use case and storing the information\\
    \textbullet & Durability
    & yes
    & with extension from this application, but also through former
    solutions \\[1ex]
    \textbullet & Immediate documentation
    & under development
    & a web portal is implemented \\[1ex]
    \textbullet & Genuineness
    & yes\newline
    customisation issue
    & combination of work flow integration in the DataFinder and the
    provenance service \\[1ex]
    \textbullet & Protocol style
    & yes \newline
    original
    & can be added as files to the system \\[1ex]
    \textbullet & Short notes
    & yes\newline
    original
    & as extra files or meta data to a data item \\[1ex]
    \textbullet & Verifying results
    & yes (rudimentary)
    & signing concept and implementation as extension \\
    \hline 
    \textbullet & Accessibility
    & yes\newline
    original
    & open source software \\[1ex]
    \textbullet & Collaboration
    & yes\newline
    original
    & same shared repository for each user, with similar information \\[1ex]
    \textbullet & Device integration
    & yes\newline
    customisation issue
    & integration via script API \\[1ex]
    \textbullet & Enabling environmental specialisation
    & yes\newline
    customisation issue
    & can be customised with scripts and data model \\[1ex]
    \textbullet & Flexible Infrastructure
    & yes\newline
    original
    & client: platform independent Python application server;\newline
    meta data: WebDAV or SVN (extensible);\newline
    data: several (extensible) \\[1ex]
    \textbullet & Individual Sorting
    & partly\newline
    under development
    & customising the view of the repositories is possible (but saving
    the settings is in planning) \\[1ex]
    \textbullet & Rights/privilege management
    & yes\newline
    under construction
    & the server supports it on the client side, the integration into
    DataFinder is currently developed \\[1ex]
    \textbullet & Variety of data formats
    & yes\newline
    original
    & any data format can be integrated, opening them depends on the
    users system \\[1ex]
    \textbullet & Searchability
    & yes\newline
    original
    & full text and meta data search \\[1ex]
    \textbullet & Versioning
    & yes
    & SVN as storage back-end is developed to enable versioned meta data
    and data \\
    \hline
  \end{tabular}
  \caption{Implementation of the laboratory notebook requirements into the
    DataFinder}
  \label{tab:DF_req_lab}
\end{table}

\subsection{Outlook: Improving DataFinder-Based Laboratory Notebook}
\label{sect:OutlookImprovingDataFinderNotebook}

Of course the systems discussed in this chapter themselves are still
research in progress and under constant development.  On the one hand
we can already envision a list of laboratory notebook features for
desired improvements.  On the other hand, this DataFinder-based
laboratory notebook can be deployed to Grid environments.

\subsubsection{Improving Laboratory Notebook Features}

After the implementation, the next step is to deploy and integrate the
system not only as a data management system, but as laboratory
notebook to suit the needs of different organisations.  For every
deployment, customisation through automation scripts and specialised
GUI dialogues need to be performed.  Particularly for the purpose of
electronic laboratory notebooks, some generic and easy integrated note
editor widget would be much appreciable for free form note taking
(instead of using external editors and importing the resulting data
files).  One a much more specialised level the following future
features are considered to be beneficial to further improve the
laboratory notebook functionality of DataFinder, and therefore meet
the requirements of other deployment scenarios:

\begin{description}
  \item[\emph{Mobile version of DataFinder:}] A mobile version of the
  data management system client could ease the scientist's
  documentation efforts when working on-site, away from the
  established (office, lab) environment.  This way the scientist could
  augment data items through notes or add/edit meta-data and data
  on-the-fly.  Requirement of immediate documentation could be met
  through this extension.

  \item[\emph{Automatic generation of reports:}] For many project
  leaders it is interesting or important to be kept up-to-date on the
  current status of a project or what their team members are currently
  doing.  For this they can currently only access the data directly.
  A feature summarising current reports and gives an intermediate
  report, could simplify the check up.  This feature was found in the
  evaluated laboratory notebook
  mbllab~\cite{mbllab-ElektronischesLaborbuch}.

  \item[\emph{Integrated standard procedures:}] In GLP, a standard
  procedures defines workflows for specific machines.  In the
  laboratory notebook mbllab~\cite{mbllab-ElektronischesLaborbuch}
  these are integrated and give the user a guideline for actions.

  \item[\emph{More elaborate signing and documenting features:}]
  Scientists discuss results of colleagues.  For more collaborative
  work situations, DataFinder needs to be enhanced with better
  features for user interactions.  On the one hand a
  discussion/commenting mechanism on data items could be supported, on
  the other hand a scientist can sign data and leave some kind of
  digital identity card.  This could be used to reference a list of
  other items signed or projects worked on.  In the evaluated
  laboratory software NoteBookMaker~\cite{NoteBookMaker}, a witness
  principle with library card is integrated.  Each notebook page
  contains an area, where a scientist can witness (authenticate) an
  entry.  After witnessing the data, the information of the witnessing
  person's identity is displayed on the corresponding page.  This
  witnessing information is then connected to a library card listing
  personal information and projects.

  \item[\emph{A graphical representation:}] A graphical representation
  of the provenance information on the server or in the DataFinder can
  help to make provenance information visually more accessible.  This
  integration of provenance data in DataFinder assists a user in
  understanding correlations between items.

  \item[\emph{Configuration options:}] Selecting a specific provenance
  or archiving system should be possible.  This could be handled
  through a new option in the data store's configuration.
  Additionally a dialog prompting for this information needs to be
  implemented.
\end{description}

\subsubsection{Migrating the Laboratory Notebook to the Grid}

Sect.~\ref{sect:MataNui} already explains how a data management system
suitable for the Grid can be constructed.  The laboratory notebook
system is ``resting'' on top of that particular data storage system,
under support of a provenance store to enable provenance enabled
working schemes.  Therefore, the two aspects of an underlying
Grid-based data storage system and of a Grid-enabled provenance store
need to be discussed.

While MongoDB with GridFS is a mature product ready to deploy, the
overlaying service infrastructure for a Grid-enabled data service is
not quite as matured.  Currently the Griffin GridFTP
server~\cite{ZhangKlossEtAl_GriffinProject} is in productive
deployment both in the Australian as well as the New Zealand eScience
infrastructures.  However the GridFS storage back-end already works,
but is still only available in a beta version and needs a little
further completion and testing.  The situation is similar with the
MataNui RESTful Web Service front end, which still needs
implementation of further query functionality.  Current tests of the
two systems have showed that throughput bottle necks to both services
currently seems mostly limited by the throughput of the underlying
disk (RAID) storage system or network interface (giga-bit ethernet),
while the database and service layer implementation is easily holding
up even on a moderately equipped system (CPU and memory).

To access this MataNui infrastructure with the DataFinder at least one
of two things still has to be implemented: The GridFTP data store
back-end needs to be ported from the 1.x line of DataFinder versions,
or a MataNui data store back-end needs to be implemented for the
current version.  For best results preferably the latter has got
priority on the list of further implementations to reach this goal.
Due to the nature of the service as well as the persistence
abstraction in the 2.x DataFinder versions, this should be relatively
straight forward.  This enables DataFinder to completely retire WebDAV
or Subversion as a centralised data server for data content as well as
meta-data, relocating this information completely onto a Grid
infrastructure.

In such a setup, DataFinder accesses the MataNui service natively,
while all managed (payload) data can be accessed through GridFTP
(Griffin server) for the purpose of compatibility with other Grid
environments.  This supports common usage for example using file
staging for Grid job submission.  Storage server side replication
ensures seamless usage in geographically distributed research teams
while retaining high throughput and low latencies through the
geographically closest storage server.

The provenance store prOOst currently does not yet support access of
its REST service through Grid authenticated means.  Once this is
implemented for the newly releases provenance store, every required
service for a Grid-enabled data service with provenance capabilities,
can be accessed using the same credentials and common Grid access
protocols.


\section{Conclusion}
\label{sect:Conclusion}

This chapter sketches a scenario of using provenance tracking with
DataFinder to support good laboratory practice and to track relations
between stored documents.  In this scenario DataFinder is used in a
distributed system together with a central provenance store.  This
makes it possible to access and update data from virtually anywhere
with a network connection, while keeping track of all interactions
with data items through recorded provenance information at any time.
When implementing the laboratory notebook, stored provenance
information can be queried to enable the extraction of additional
valuable meta-data information on data items.  As a result, provenance
is successfully used to trace typical scientific workflows comprising
of preparation, execution, evaluation, interpretation and archiving of
research data.  The reliability -- and therefore credibility -- of
research results is increased, and assistance to help understand
involved processes is provided for the researcher.

Such a system can be implemented on top of a Grid data infrastructure,
as the described MataNui system.  The MataNui service is mostly
functioning already, but still needs integration into DataFinder as a
full-featured storage back-end for data as well as meta-data.
Additionally, it is already possible to expose the data repository to
Grid environments directly using the GridFTP protocol.  GridFTP is
commonly used for scripts, automation and compatibility with other
Grid enabled tools.  The overall MataNui concept has been designed to
be capable of handling files large in number and size, as well as
manage arbitrary amounts of meta-data associated with each data item.
It is usable in distributed projects with a self-replicating,
federated data infrastructure.  This federation can drastically
improve data access latency and throughput by connecting to a
geographically close service.  Through support for server side
queries, meta-data searches can be processed very efficiently by
avoiding transfers of potentially large numbers of data sets to a
client.  Lastly, the implementation of MataNui has been undertaken
with the vision of it being robust as well as easy to deploy and use.


\pagebreak
\bibliographystyle{spmpsci}
\bibliography{references}
\end{document}